# A multivariate extension of the Lorenz curve based on copulas and a related multivariate Gini coefficient

**Oliver Grothe[1] · Fabian Kächele[1]**  **· Friedrich Schmid[2]**

## Abstract

We propose an extension of the univariate Lorenz curve and of the Gini coefficient to the multivariate case, i.e., to simultaneously measure inequality in more than one variable. Our extensions are based on copulas and measure inequality stemming from inequality in each single variable as well as inequality stemming from the dependence structure of the variables. We derive simple nonparametric estimators for both instruments and exemplary apply them to data of individual income and wealth for various countries.

**Keywords** Copula · Gini coefficient · Inequality · Multidimensional inequality · Joint inequality of income and wealth · Lorenz curve · Dependence of income and wealth

**Abbreviations**
MEGC    Multivariate extension of the Gini coefficient
MEILC   Multivariate extension of the inverse Lorenz curve

## 1 Introduction

The well known Lorenz curve and Gini coefficient are still the most important tools for representation and analysis of inequality in a distribution, such as the income and wealth distribution. Both, however, are univariate instruments, i.e., they analyze the variables individually, ignoring their dependence structure. Considering the example of income and

✉ Fabian Kächele
   fabian.kaechele@kit.edu

   Oliver Grothe
   oliver.grothe@kit.edu

1   IOR - Analytics and Statistics, Karlsruher Institut für Technologie (KIT), Kaiserstraße 12, 76131 Karlsruhe, Germany

2   Institut für Ökonometrie und Statistik, Universität zu Köln, Albertus-Magnus-Platz, D-50923 Köln, Germany

 Springer

wealth, it is not possible to see the differences in the overall inequality if wealthy people coincide with high-income people compared to a more balanced, eventually compensating distribution of wealth over the income groups. Contrary to that, in this paper, we propose extensions of both tools to study the inequality of $d$ variables $X_1, \ldots, X_d$ simultaneously. By that, we explicitly capture the dependence structure of these variables which gets lost if only one variable is considered at a time.

There had been some efforts to extend Lorenz curve and Gini coefficient to the multivariate case before. The earliest suggestion in this direction we know of is Taguchia (1972a, b) who applied methods of differential geometry. Further suggestions are by Arnold (1987), Arnold and Sarabia (2018), Gajdos and Weymark (2005) and Koshevoy and Mosler (1996, 1997). We will not give an overview of these contributions because this is - at least partially - done by Arnold and Sarabia (2018). We agree with the view of the latter authors that all extensions are essentially determined by (elegant) mathematical considerations but may lack interpretability and economic reasoning.

Here, we propose direct and natural multivariate extensions of both, the (inverse) Lorenz curve and the Gini coefficient. We exploit the fact that the inverse of a variable's Lorenz curve is the distribution function of a simple monotonically increasing transformation of that variable. The multivariate inverse Lorenz curve of the $d$ variables $X_1, \ldots, X_d$ is then defined as the joint distribution function of analogous univariate transformations of them. The resulting Lorenz curve can explicitly be expressed using copulas. Copulas decompose the joint distribution function of variables into marginal distribution functions and their dependence structure. Consequently, for a given vector $X = (X_1, \ldots, X_d)$ of $d$ variables, the copula based multivariate Lorenz curve identifies and captures two different sources of inequality:

a. inequality contained in the individual variables $X_i$, measured by the univariate Lorenz curve $L_i$ or inverse Lorenz curve $L_i^{-1}$ for $i = 1, \ldots, d$.
b. inequality due to the dependence structure of the variables $X_1, \ldots, X_d$ which is captured by the copula of these variables.

Based on the multivariate Lorenz curve, the formulation of a multivariate Gini coefficient follows in a natural way analogously to the derivation of the univariate Gini coefficient from the univariate Lorenz curve.

The mathematics we apply in the theoretical part of the paper is some elementary copula theory and - hopefully - is accessible to a broad readership. Later on in the paper, we derive simple nonparametric estimators for both instruments, and provide ready to use computer code in the supplementary material of this paper. We illustrate both instruments, the multivariate Lorenz curve and Gini coefficient, on data sets consisting of individual wealth and income data of various countries. The results are intuitive and show that the two above mentioned sources a. and b. of multivariate inequality are reflected in a reasonable and interpretable way.

The structure of the paper is the following. Section 2 introduces notation and some definitions. The multivariate extension of the inverse Lorenz curve (MEILC) is introduced in Section 3. Various properties of the MEILC are derived. Section 4 presents a multivariate extension of the Gini coefficient (MEGC) related to the MEILC and considers the bivariate Gini (i.e., $d = 2$) as a special case. Nonparametric estimation of MEILC and MEGC is considered in Section 5. In Section 6 we address some aspects regarding multivariate transfers. The last section of the paper contains the empirical applications.

## 2 Notations and definitions

We consider a $d$-variate vector of random variables $X = (X_1, \ldots, X_d)$ defined on a probability space $(\Omega, \Sigma, \mathbb{P})$. The joint distribution function is given by $F_X(x) = P(X_1 \leq x_1, \ldots, X_d \leq x_d)$ for $x = (x_1, \ldots, x_d) \in \mathbb{R}^d$ and the marginal distribution functions $F_i$ of $X_i$ are given by $F_i(x_i) = P(X_i \leq x_i)$ for $x_i \in \mathbb{R}$ and $i = 1, \ldots, d$. Throughout this paper we assume that $X_i \geq 0$ and $0 < \mu_i = E(X_i) < \infty$ for $i = 1, .., d$. Note that all variables should fit together in an economically meaningful way and should have a cardinal scale. For inequality measurement in the case of ordinal or qualitative data we refer the interested reader to Allison and Foster (2004) and Gravel et al. (2021) or Kobus and Miłoś (2012) among others.

There exist a copula $C = C_{F_X}$, such that $F_X(x_1, .., x_d) = C(F_1(x_1), \ldots, F_d(x_d))$ for $(x_1, \ldots, x_d) \in \mathbb{R}^d$. The copula $C$ is uniquely determined if the marginal distribution functions $F_i$ are continous. The basics of copulas are given in Nelson (2006), Joe (2014), and Durante and Sempi (2015) and also have been applied in the context in income analysis by Aaberge et al. (2018) recently. An important property is that the copula $C$ of a random vector $X$ is invariant with respect to strictly increasing transformations of the marginal distributions.

Every copula $C$ is for any $u = (u_1, \ldots, u_d) \in [0, 1]^d$ bounded by the Fréchet–Hoeffding bounds:

$$W(u_1, \ldots, u_d) \leq C(u_1, \ldots, u_d) \leq M(u_1, \ldots, u_d), \quad (1)$$

with $W(u_1, \ldots, u_d) = \max\left\{1 - d + \sum_{i=1}^{d} u_i, 0\right\}$ and $M(u_1, \ldots, u_d) = \min\{u_1, \ldots, u_d\}$. The upper bound $M$, which is called comonotonicity copula, corresponds to the dependence structure of full monotone positive dependence. An example of wealth and income with such a dependence structure would be a population of size $N$, where the $i-$th wealthiest individual has also the $i-$th highest income (for $i = 1, \ldots, N$). The lower bound $W$ is only a proper copula in the bivariate case and then called the countermonotonicity copula. In a countermonotonic income/wealth example, the $i-$th wealthiest individual would have the $i-$th lowest income (for $i = 1, \ldots, N$). If $X$ is a vector of independent variables $X_1, \ldots, X_d$ the corresponding copula is the independence copula $\Pi$ with $\Pi(u_1, \ldots, u_d) = \prod_{i=1}^{d} u_i$. There is a partial order $C \leq C'$ on the set of d-variate copulas given by $C(\mathbf{u}) \leq C'(\mathbf{u})$ for all $\mathbf{u} \in [0, 1]^d$ (see Nelson 2006).

The univariate Lorenz curve for $i = 1, \ldots, d$ is given by

$$L_i(u_i) = \frac{1}{\mu_i} \int_0^{u_i} F_i^{-1}(t) dt \text{ for } u_i \in [0, 1] \text{ and in } i = 1, \ldots, d, \quad (2)$$

see Gastwirth (1971). Each $L_i$ is a continuous, weakly increasing and weakly convex function. It has all the properties of a distribution function if we extend $L_i$ by 1 for $u_i > 1$ and by 0 for $u_i < 0$. The inverse of $L_i$ is defined by

$$L_i^{-1}(u_i) = \begin{cases} inf\{t | L_i(t) \geq u_i\}, & \text{for } u_i \in \, ]0, 1] \\ sup\{t | L_i(t) = u_i\}, & \text{for } u_i = 0 \end{cases}$$

and $L_i^{-1}$ is continuous, weakly increasing and weakly concave on $[0, 1]$. It has all the properties of a distribution function if we extend $L_i^{-1}$ by 1 for $u_i > 1$ and by 0 for $u_i < 0$. Note that there might be a point mass at zero.

Consider individual income in a population. The usual interpretation of the Lorenz curve $L$ of this variable is that, e.g., for $p \in [0, 1]$, $L(p)$ denotes the proportion of total income that corresponds to the bottom $p \cdot 100\%$ of the individuals. The interpretation of the inverse Lorenz curve is that, for $q \in [0, 1]$, $L_i^{-1}(q)$ indicates the maximum percentage of the population with a combined cumulative share of $q \cdot 100\%$ of the total income (the maximum ensures starting with the bottom income individual here). Obviously, both curves describe the inequality in an equivalent way. It is worth mentioning that Lorenz (1905) originally proposed in his paper what we now call the inverse Lorenz curve.

Using $X_i$, $F_i$ and $L_i$ as defined above we now define the following random variables $X_i^*$ by

$$X_i^* = L_i(F_i(X_i)), \text{ for } i = 1, \ldots, d. \tag{3}$$

Note, the difference between $X_i$ and $X_i^*$. In applications $X_i$ has a dimension (such as income or wealth). $X_i^*$, however, is a fraction (i.e., a number between 0 and 1). If, e.g., $X_i$ denotes again individual income in a population then $X_i^*$ is the corresponding joint fraction of the total income of that part of the population having individual incomes smaller or equal to $X_i$. The $d$-variate vector $\boldsymbol{X}^*$ is defined by $\boldsymbol{X}^* = (X_1^*, \ldots, X_d^*)$. The marginal distribution function for $X_i^*$ is given by the inverse Lorenz curve of $X_i$, i.e.,

$$F_{X_i^*}(u_i) = P(L_i(F_i(X_i)) \leq u_i) = L_i^{-1}(u_i)$$

for $u_i \in [0, 1]$ and $i = 1, \ldots, d$. The joint distribution function of $\boldsymbol{X}^*$ is given by

$$F_{\boldsymbol{X}^*}(u_1, \ldots, u_d) = P(X_1^* \leq u_1, \ldots, X_d^* \leq u_d) \tag{4}$$

$$= C\left( L_1^{-1}(u_1), \ldots, L_d^{-1}(u_d) \right) \text{ for } u_i \in [0, 1] \text{ and } i = 1, \ldots, d. \tag{5}$$

Note, that the copula of $\boldsymbol{X}$ is identical to the copula of $\boldsymbol{X}^*$, since $X_i^*$ is a monotonically increasing function of $X_i$ for $i = 1, \ldots, d$.

The univariate Gini coefficient is defined as a normalization of the area enclosed by the Lorenz curve and the diagonal of the unit square. It equals one minus twice the area under the Lorenz curve (Kakwani 1977; Gastwirth 1972)

$$G = 1 - 2 \int_{[0,1]} L(u)du.$$

Considering that $1 - \int L(u)du = \int L^{-1}(u)du$ and $L^{-1}$ is the $cdf$ of $X^*$, it follows that $\int L(u)du = E(X^*)$ and the univariate Gini coefficient may be expressed as

$$G = 1 - 2E(X^*)$$

as well as

$$G = 2 \int_{[0,1]} L^{-1}(u)du - 1$$

when using the inverse Lorenz curve $L^{-1}$ and considerations above.

Notation and definitions introduced in this section are used to define a multivariate extension of the univariate Lorenz curve (see Section 3) and a multivariate extension of the univariate Gini coefficient (see Section 4).

## 3 A multivariate extension of the Lorenz curve based on copulas (MEILC)

As mentioned in the introduction, inequality in a $d$-variate random vector $\boldsymbol{X} = (X_1, \ldots, X_d)$ has two different sources:

a. inequality in the individual variables $X_i$, which is measured by the corresponding Lorenz curves $L_i(u_i)$ or inverse Lorenz curves $L_i^{-1}(u_i)$ for $i = 1, \ldots, d$ and $u_i \in [0, 1]$.
b. inequality contained in the dependence structure of the vector $\boldsymbol{X} = (X_1, \ldots, X_d)$ which is represented by the copula $C$ of $\boldsymbol{X}$.

To illustrate the effect of b. on the joint inequality in $\boldsymbol{X} = (X_1, \ldots, X_d)$ in more detail, we look at a very simple example for the bivariate case, i.e., $d = 2$, and a population of five individuals, where $X_1$ and $X_2$ might again stand for individual income and wealth, respectively.

|  | | Individual 1 | Individual 2 | Individual 3 | Individual 4 | Individual 5 |
|---|---|---|---|---|---|---|
| Society 1 | $\begin{pmatrix} X_1 \\ X_2 \end{pmatrix}$ | $\begin{pmatrix} 1 \\ 1 \end{pmatrix}$ | $\begin{pmatrix} 2 \\ 2 \end{pmatrix}$ | $\begin{pmatrix} 3 \\ 3 \end{pmatrix}$ | $\begin{pmatrix} 4 \\ 4 \end{pmatrix}$ | $\begin{pmatrix} 5 \\ 5 \end{pmatrix}$ |
| Society 2 | $\begin{pmatrix} X_1 \\ X_2 \end{pmatrix}$ | $\begin{pmatrix} 1 \\ 3 \end{pmatrix}$ | $\begin{pmatrix} 2 \\ 2 \end{pmatrix}$ | $\begin{pmatrix} 3 \\ 5 \end{pmatrix}$ | $\begin{pmatrix} 4 \\ 1 \end{pmatrix}$ | $\begin{pmatrix} 5 \\ 4 \end{pmatrix}$ |
| Society 3 | $\begin{pmatrix} X_1 \\ X_2 \end{pmatrix}$ | $\begin{pmatrix} 1 \\ 5 \end{pmatrix}$ | $\begin{pmatrix} 2 \\ 4 \end{pmatrix}$ | $\begin{pmatrix} 3 \\ 3 \end{pmatrix}$ | $\begin{pmatrix} 4 \\ 2 \end{pmatrix}$ | $\begin{pmatrix} 5 \\ 1 \end{pmatrix}$ |

It can be seen that the marginal distributions of $X_1$ and $X_2$ over the five individuals are the same in these societies. We think that it is quite obvious to see that the inequality is largest in Society 1 and smallest in Society 3. Society 2 is somewhere in between.

The differences in joint inequality in these societies are due to different dependence structures between the variables. In terms of copulas, the dependence structure in Society 1 corresponds to the comonotoncity copula $M$, the upper bound in the set of bivariate copulas. In contrast, Society 3 corresponds to the countermonotonicity copula $W$ which is the lower bound in the set of bivariate copulas. Thus, in Society 1, the high income individuals are also the wealthiest, whereas in Society 3, income and wealth kind of compensate each other. Society 2 might stem from the independence copula $\Pi$. We conclude from this example that joint inequality in a vector $\boldsymbol{X} = (X_1, \ldots, X_d)$ is increasing in C in the partial order as defined in Section 2.

Having the example in mind, we now define a multivariate extension of univariate (inverse) Lorenz curves considering the dependence structure of the variables. It will turn out that the multivariate Lorenz curve of random vector $\boldsymbol{X}$ is the joint distribution function (compare to Eq. 4) of the random vector $\boldsymbol{X^*}$ as defined in Eq. 3 in Section 2.

**Definition 3.1** Multivariate extension of the inverse Lorenz curve (MEILC) and Lorenz order
Using the notation of Section 2, let

1. $\mathbb{L}_{C,L_1^{-1},\ldots,L_d^{-1}}^{-1}(u_1, \ldots, u_d) = C\left(L_1^{-1}(u_1), \ldots, L_d^{-1}(u_d)\right)$ for $(u_1, \ldots, u_d) \in [0, 1]^d$.

2.  For a second vector $\tilde{X} = (\tilde{X}_1, \ldots, \tilde{X}_d)$ with copula $\tilde{C}$ and inverse Lorenz curves $\tilde{L}_i^{-1}(u_i)$ of $\tilde{X}_i$ for $i = 1, \ldots, d$, we define the multivariate ordering $\tilde{X} \succeq X$ if and only if
    $$\mathbb{L}_{\tilde{C}, \tilde{L}_1^{-1}, \ldots, \tilde{L}_d^{-1}}^{-1}(\boldsymbol{u}) \geq \mathbb{L}_{C, L_1^{-1}, \ldots, L_d^{-1}}^{-1}(\boldsymbol{u}) \text{ for all } \boldsymbol{u} = (u_1, \ldots, u_d) \in [0,1]^d.$$

The extension of the inverse Lorenz curve (MEILC) $\mathbb{L}^{-1}(\boldsymbol{u}) = \mathbb{L}_{C, L_1^{-1}, \ldots, L_d^{-1}}^{-1}(\boldsymbol{u})$ has a nice interpretation in terms of the $X_i^*$ for $i = 1, \ldots, d$ and $\boldsymbol{u} = (u_1, \ldots, u_d) \in [0,1]^d$. Since $\mathbb{L}^{-1}(u_1, \ldots, u_d)$ is the joint distribution function of $X^* = (X_1^*, \ldots, X_d^*)$ we see that $\mathbb{L}^{-1}(u_1, \ldots, u_d)$ is the population fraction for which $X_1^* \leq u_1, \ldots, X_d^* \leq u_d$ and therefore the fraction with a cumulative share of the features smaller or equal to $u_1, \ldots, u_d$. E.g., for $d = 2$ if $u_1$ denotes a share of the cumulative income in a population and $u_2$ a share of wealth, than $\mathbb{L}^{-1}(u_1, u_2)$ is the corresponding fraction of people collectively having not more than shares $u_1$ and $u_2$ of the total income and wealth, respectively. Note that the interpetation of the MEILC coincides with the interpretation of the upper parts of the *Lorenz zonoid* introduced by Koshevoy and Mosler ([1996](#)). However, while calculating zonoids from data is computationally intensive, the copula approach results in simple formulas, also allowing a straight forward extension of the Gini coefficient later in the paper.

We analyze some of the (formal) properties of the MEILC. Here, we are in particular interested in how $\mathbb{L}^{-1}(u_1, \ldots, u_d)$ behaves, when ceterus paribus either marginal inequalities or the dependence structures are changed. Later in the paper, in Section [6](#), we discuss some implied properties such as the reaction to transfers in empirical data.

1.  Obviously $\mathbb{L}^{-1}(u_1, \ldots, u_d)$ is a function from $[0,1]^d$ to $[0,1]$. Furthermore for every $C, L_1^{-1}(u_1), \ldots, L_d^{-1}(u_d)$ and $\boldsymbol{u} = (u_1, \ldots, u_d) \in [0,1]^d$ we have
    $$\mathbb{L}_{min}^{-1}(\boldsymbol{u}) \leq \mathbb{L}_{C, L_1^{-1}, \ldots, L_d^{-1}}^{-1}(\boldsymbol{u}) \leq \mathbb{L}_{max}^{-1}(\boldsymbol{u})$$
    where
    $$\mathbb{L}_{min}^{-1}(\boldsymbol{u}) = W(u_1, \ldots, u_d) = max\{0, \textstyle\sum_i^d u_i - (d-1)\}$$
    and
    $$\mathbb{L}_{max}^{-1}(\boldsymbol{u}) = M(1, \ldots, 1) = min\{1, \ldots, 1\} = 1.$$
    These boundaries follow directly from the Fréchet–Hoeffding bounds. Regarding the margins, note that the arguments of the lower bound refer to minimal marginal inequality, i.e., $L_i^{-1}(u_i) = u_i$, for $i = 1 \ldots d$, whereas the arguments of the upper bound refer to maximal marginal inequality, i.e., $L_i^{-1}(u_i) \equiv 1$, for $i = 1 \ldots d$. Thus, e.g., the upper bound corresponds to the case of maximal marginal inequality as well as maximal dependence between the variables and reflects thus the case of maximal multivariate inequality.

2.  If $X_1, \ldots, X_d$ are independent, i.e., $C = \Pi$ we have
    $$\mathbb{L}_{\Pi, L_1^{-1}, \ldots, L_d^{-1}}^{-1}(\boldsymbol{u}) = \textstyle\prod_{i=1}^d L_i^{-1}(u_i) \text{ for } \boldsymbol{u} = (u_1, \ldots, u_d) \in [0,1]^d.$$
    So the MEILC is the product of the univariate Lorenz curves in such cases.

3.  If $u_d = 1$ we know
    $$\mathbb{L}_{C, L_1^{-1}, \ldots, L_{d-1}^{-1}, L_d^{-1}}^{-1}(u_1, \ldots, u_{d-1}, 1) = \mathbb{L}_{C, L_1^{-1}, \ldots, L_{d-1}^{-1}}^{-1}(u_1, \ldots, u_{d-1}) \text{ for}$$
    $(u_1, \ldots, u_{d-1}) \in [0,1]^{d-1}$ and similar formulas hold for $i = 1, \ldots, d-1$ as well as for more general index sets $I \subset \{1, \ldots, d\}$.
    This marginalization is quite intuitive, since setting $u_i = 1$ in the MEILC refers to the fraction of the population having less or equal than the maximum value of $X_i$.

Therefore, the $i$-th dimension is not restrictive anymore, while the other dimensions still are.

By further marginalizing, we can see that for $d = 1$, we get the univariate (inverse) Lorenz curve as a margin, e.g., $\mathbb{L}^{-1}_{C, L_1^{-1}}(u_1) = L_1^{-1}(u_1)$ for $u_1 \in [0, 1]$.

4. If for $\boldsymbol{u} = (u_1, \ldots, u_d) \in [0, 1]^d$ at least one $L_i^{-1}(u_i)$ is zero, then $\mathbb{L}^{-1}(\boldsymbol{u})$ is zero. But note that there might be point masses at zero in some or even all of the $X_i$. The MEILC therefore does not necessarily start at zero since a point mass of $X_i$ at zero would imply $L_i^{-1}(0) > 0$.

5. Response of $\mathbb{L}^{-1}(\boldsymbol{u})$ to changing $L_i^{-1}(u_i)$ for fixed $\boldsymbol{u} = (u_1, \ldots, u_d) \in [0, 1]^d$: Higher values of $L_i^{-1}(u_i)$ lead to higher values of $\mathbb{L}^{-1}(\boldsymbol{u})$, ceteris paribus. This follows directly from the definition of $\mathbb{L}^{-1}(\boldsymbol{u})$ and general properties of every copula. An increased inequality in one dimension therefore leads to an increased total inequality without any further changes.

6. Response of $\mathbb{L}^{-1}(\boldsymbol{u})$ to changes in the dependence structure of the variables, i.e., to changes of $C$, when the $L_i^{-1}(u_i)$ do not change: Consider two copulas $C_A$ and $C_B$ with $C_A(\boldsymbol{u}) \leq C_B(\boldsymbol{u})$ for all $\boldsymbol{u} \in [0, 1]^d$. Here, referring to the example of income and wealth, in B the wealthy would tend to belong more to the high-income part of the society than in A. It then follows the corresponding multivariate Lorenz order from Definition 3.1, i.e., $\mathbb{L}^{-1}_A(\boldsymbol{u}) \leq \mathbb{L}^{-1}_B(\boldsymbol{u})$. Generally, the order properties of the involved copulas transfer directly to the multivariate Lorenz order. Since the copula order is, however, a partial order, not all changes in the dependence structure lead to ordered Lorenz curves.

We illustrate the MEILC for some bivariate examples in Figs. 1 and 2. We consider two different types of marginal Lorenz curves in all examples, $L_i(u_i) = u_i^2$ (this corresponds to values $X_i$ which are uniformly distributed over a finite interval $[0, b]$ with $b > 0$) or $L_i(u_i) = u_i^{10/9}$ (which is close to the Lorenz curve of minimal inequality). Note that the corresponding marginal inverse Lorenz curves are $L_i^{-1}(u_i) = \sqrt{u_i}$ and $L_i^{-1}(u_i) = u_i^{0.9}$, respectively. In Fig. 1, we consider Gaussian dependence structures of the variables and vary Spearman's $\rho$ from negative to positive dependence, starting from the case of strong negative dependence (a) to the case of independent margins (b), small positive dependence (c) and strong positive dependence (d). As expected, the surface of the MEILC becomes more domed for increasing strength of dependence. Recall that a point on the surface $\mathbb{L}^{-1}(u_1, u_2)$ at $(u_1, u_2)$ reflects the maximum share of the society having together less than shares $u_1$ and $u_2$ of the total variable sums of $X_1$ and $X_2$, respectively. Thus, it refers to the share of individuals being at the bottom in both variables. A more domed surface therefore reflects a larger inequality.

In Fig. 2, we illustrate the effect of the copula family and a case of unequal marginal inverse Lorenz curves. Panels (a) and (b) both refer to cases with a rank correlation of $X_1$ and $X_2$ of $\rho = 0.8$ but different asymmetric dependence structures, i.e., copulas. The Clayton copula (a) has a stronger dependence between small values, while the Gumbel copula (b) has strongest dependence between large values. Consequently, we see that the surface of the MEILC in the Clayton case is more domed for pairs of small values than in the Gumbel case (b). Panel (c) depicts the case of independence where the margins are now different. It can be seen that surface interpolates between the margins. Again, the surface gets more domed, if the dependence is increased, e.g., by using a Clayton copula with $\rho = 0.8$ (d). This is done by visualising $\mathbb{L}^{-1}(\boldsymbol{u})$ for data with Gaussian and Archimedean

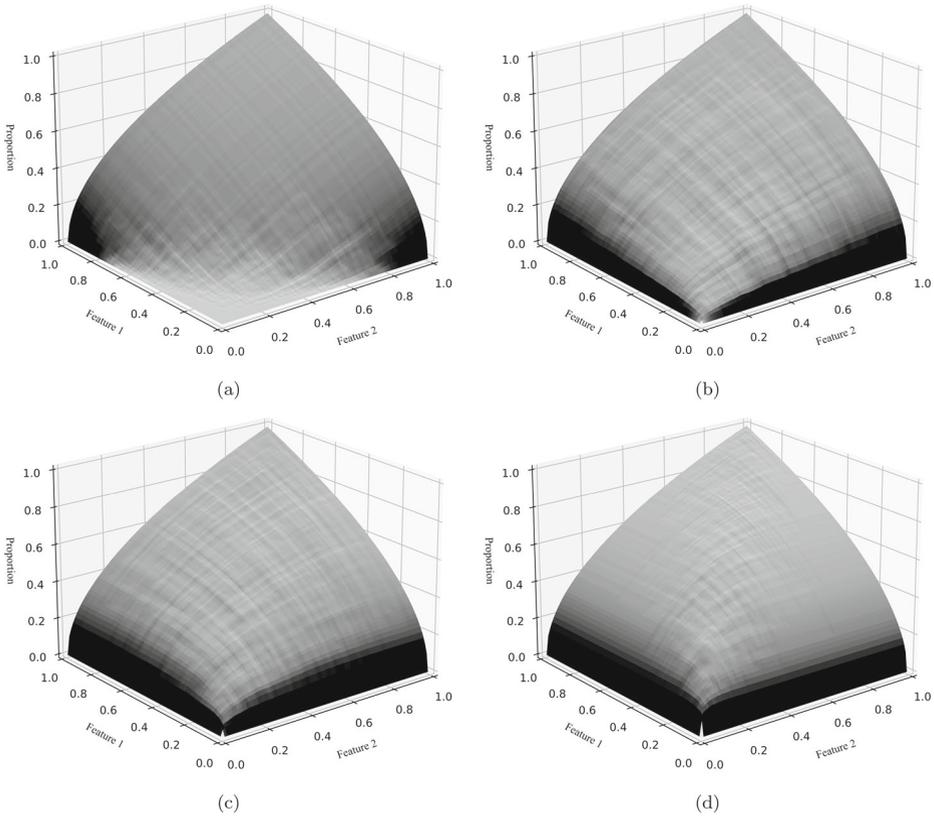

**Fig. 1** Graphs of $\mathbb{L}^{-1}(u_1, u_2)$ based on Gaussian copulas with different values for the dependence parameter Spearman's $\rho$. The surface of $\mathbb{L}^{-1}(u_1, u_2)$ gets more domed with increasing parameter $\rho$. **a** Gaussian Copula, Spearman's $\rho = -0.9$ and $L_i^{-1}(u_i) = \sqrt{u_i}$ for $i = 1, 2$. **b** Gaussian Copula, Spearman's $\rho = 0.0$ (equals the independence copula) and $L_i^{-1}(u_i) = \sqrt{u_i}$ for $i = 1, 2$. **c** Gaussian Copula, Spearman's $\rho = 0.5$ and $L_i^{-1}(u_i) = \sqrt{u_i}$ for $i = 1, 2$. **d** Gaussian Copula, Spearman's $\rho = 0.9$ and $L_i^{-1}(u_i) = \sqrt{u_i}$ for $i = 1, 2$

copulas (see Nelson 2006 Chapter 2-4 for various parametric copulas) as well as different dependence parameter Spearmann's $\rho$ and marginal distributions.

*Remark 3.1* It might be surprising that our extension of the Lorenz curve is based on its inverse and not on the Lorenz curve itself. The inverse Lorenz curve draws proportions of the people on the y-axis and the variable of interest, e.g., share of total income, on the x-axis. Proportion of people is thus the value of the inverse function, while the variable of interest is the argument. Having only one variable of interest, the choice between Lorenz curve or inverse Lorenz curve seems arbitrary. Considering $d > 1$ variables of interest, however, it seems conceptually more natural to add these variables as further arguments of the inverse Lorenz curve. Furthermore, the resulting extension is easily interpretable.

Alternatively, starting from the Lorenz curve, a seemingly intuitive idea like

$$(u_1, \ldots, u_d) \longmapsto C(L_1(u_1), \ldots, L_d(u_d)).$$

behaves contradictorily. If inequality in the $X_i$ rises than the above definition indicates a decreasing value. If inequality accounted in C increases an increasing value is indicated.

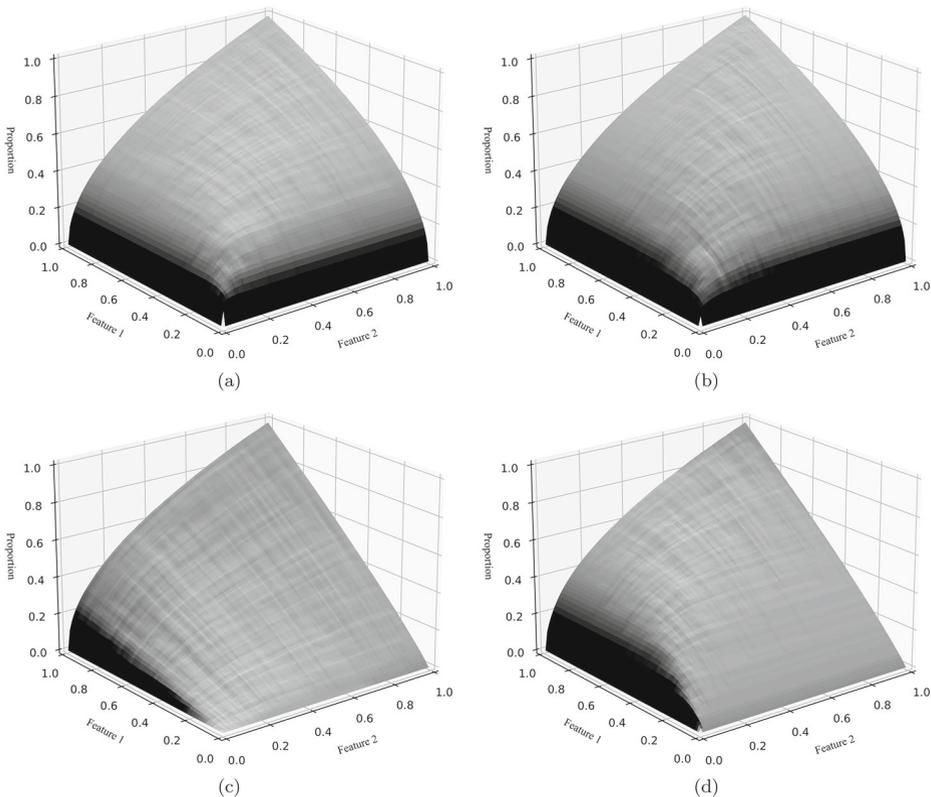

**Fig. 2** Graphs of $\mathbb{L}^{-1}(u_1, u_2)$ based on Archimedean copulas with different values for the dependence parameter Spearman's $\rho$ and marginal distributions. Panels (a) and (b) illustrate the effect of different asymmetric dependence structures, while (c) and (d) illustrate effects of margins and dependence structure. **a** Clayton copula, Spearman's $\rho = 0.8$ and $L_i^{-1}(u_i) = \sqrt{u_i}$ for $i = 1, 2$. **b** Gumbel copula, Spearman's $\rho = 0.8$ and $L_i^{-1}(u_i) = \sqrt{u_i}$ for $i = 1, 2$. **c** Independence copula $\Pi$ and $L_1^{-1}(u_1) = u_1^{0.9}$ and and $L_2^{-1}(u_2) = \sqrt{u_2}$. **d** Clayton copula, Spearman's $\rho = 0.8$ and $L_1^{-1}(u_1) = u_1^{0.9}$ and $L_2^{-1}(u_2) = \sqrt{u_2}$

Obviously this is contradictory.

A possible adjustment would be to look at

$$(u_1, \ldots, u_d) \longmapsto 1 - C(1 - L_1(u_1), \ldots, 1 - L_d(u_d))$$

This object reacts in the expected directions in all cases, but it lacks a reasonable and convincing interpretation.

*Remark 3.2* We are quite aware that there might be further reasonable ways of combining a copula with Lorenz curves $L_i$ or its inverses. The survival copula $\overline{C}$ which corresponds to copula $C$ (see Nelson 2006 p.33) might also be a useful tool for the definition of a multivariate extension of the Lorenz curve, but we have not derived any details.

## 4 A multivariate extension of the Gini coefficient (MEGC) related to the multivariate extension of the Lorenz curve

In the univariate setting, i.e., $d = 1$, it is well known that the Lorenz order is only a weak order. Indeed, Lorenz curves may intersect and consequently $X_1$ and $\tilde{X}_1$ with intersecting Lorenz curves cannot be ordered with respect to inequality. A numerical measure of inequality, such as the Gini coefficient, is called for.

In the multivariate setting it can be seen that the order defined in Section 4 is a also a weak order and a related numerical measure of inequality is required, too. Using the notation of Sections 2 and 3 we define a Gini coefficient related to the MEILC as follows.

**Definition 4.1** Multivariate extension of the Gini coefficient (MEGC)

The MEGC is defined as:

$$G_{C,L_1^{-1},\dots,L_d^{-1}} = \frac{\int_{[0,1]^d} \mathbb{L}^{-1}_{C,L_1^{-1},\dots,L_d^{-1}}(\boldsymbol{u})d\boldsymbol{u} - \int_{[0,1]^d} \mathbb{L}^{-1}_{min}(\boldsymbol{u})d\boldsymbol{u}}{\int_{[0,1]^d} \mathbb{L}^{-1}_{max}(\boldsymbol{u})d\boldsymbol{u} - \int_{[0,1]^d} \mathbb{L}^{-1}_{min}(\boldsymbol{u})d\boldsymbol{u}}. \tag{6}$$

Using $\int_{[0,1]^d} \mathbb{L}^{-1}_{max}(\boldsymbol{u})d\boldsymbol{u} = 1$ and $\int_{[0,1]^d} \mathbb{L}^{-1}_{min}(\boldsymbol{u})d\boldsymbol{u} = \frac{1}{(d+1)!}$ (see Nelson 2006) we can rewrite,

$$G_{C,L_1^{-1},\dots,L_d^{-1}} = \frac{(d+1)! \int_{[0,1]^d} \mathbb{L}^{-1}_{C,L_1^{-1},\dots,L_d^{-1}}(\boldsymbol{u})d\boldsymbol{u} - 1}{(d+1)! - 1}.$$

Note, the similarity of the above definition to the univariate Gini coefficient. The latter is two times the area between the inverse Lorenz curve and the diagonal of the unit square, where the diagonal stands for the inverse Lorenz curve of minimal inequality. The factor two results from normalization to the unit interval. In our the multivariate definition, we measure the volume enclosed by the actual Lorenz curve and the curve of minimal inequality (see numerator of Eq. 6) and rescale the result to be between 0 and 1 (see denominator of that equation). Consequently, setting $d = 1$ yields,

$$G_{C,L_1^{-1}} = 2 \int_{[0,1]} L_1^{-1}(u_1)du_1 - 1 = G_1$$

which is the Gini coefficient in the univariate case (see, e.g., Section 2).

For $d = 2$ we have $0 \leq G_{C,L_1^{-1},L_2^{-1}} \leq 1$, where $G_{C,L_1^{-1},L_2^{-1}} = 0$ implies that $C = W$ and $G_{C,L_1^{-1},L_2^{-1}} = 1$ implies that $C = M$. For $d \geq 3$ we have $0 < G_{C,L_1^{-1},\dots,L_d^{-1}} \leq 1$. This is due to the fact, that $W$ is not a copula for $d \geq 3$. Further $G_{C,L_1^{-1},\dots,L_d^{-1}} = 1$ implies that $C = M$.

*Example 4.1* Consider the case of independent $X_1, \dots, X_d$, where $cov(X_i, X_j) = 0$, for $i, j \in \{1, \dots, d\}$ and $i \neq j$. With $C = \Pi$ and $\boldsymbol{u} = (u_1, \dots, u_d)$ it follows from

$$\int_{[0,1]^d} \mathbb{L}^{-1}_{\Pi,L_1^{-1},\dots,L_d^{-1}}(\boldsymbol{u})d\boldsymbol{u} = \left(1 - E(X_1^*)\right)\left(1 - E(X_2^*)\right)\dots\left(1 - E(X_d^*)\right)$$

$$= \left(\frac{1}{2}\right)^d \prod_{j=1}^d (1 + G_j)$$

and Definition 4.1 that the MEGC can be written as

$$G_{C,L_1^{-1},\dots,L_d^{-1}} = \frac{(1+d)!(\frac{1}{2})^d \prod_{j=1}^d (1+G_j) - 1}{(d+1)! - 1}.$$

In the special case $d = 2$, we obtain

$$G_{1,2} = \frac{1}{10} (1 + 3G_1 + 3G_2 + 3G_1 G_2).$$ (7)

Focusing on the two dimensional case, the MEGC is decomposable into the marginal Gini coefficients and a term resulting from the dependence structure.

**Theorem 4.1** *Decomposition of two dimensional MEGC*
*The two dimensional MEGC can be expressed as*

$$G_{C, L_1^{-1}, L_2^{-1}} = G_{1,2} = \frac{6}{5} E(X_1^* X_2^*) + \frac{3}{5} G_1 + \frac{3}{5} G_2 - \frac{1}{5}.$$

*Proof* Note, that $cov(X_1^*, X_2^*) = \int_0^1 \int_0^1 \Big( C(L_1^{-1}(u_1), L_2^{-1}(u_2)) - L_1^{-1}(u_1) L_2^{-1}(u_2) \Big) du_1 du_2$
(Nelson 2006) and remember that the $cdf$ of $X_i^*$ is $L_i^{-1}$, for $i = 1, 2$. So $\int L_1^{-1}(u_1) du_1 = 1 - \int L_1(u_1) du_1$, with $\int L_1(u_1) du_1 = E(X_1^*)$. It follows

$$\int_{[0,1]^2} \mathbb{L}_{C, L_1^{-1}, L_2^{-1}}^{-1}(u_1, u_2) du_1 du_2 = cov(X_1^*, X_2^*) + \int_0^1 L_1^{-1}(u_1) du_1 \int_0^1 L_2^{-1}(u_2) du_2$$

$$= cov(X_1^*, X_2^*) + (1 - E(X_1^*))(1 - E(X_2^*))$$

$$= cov(X_1^*, X_2^*) + 1 - E(X_1^*) - E(X_2^*) + E(X_1^*)E(X_2^*)$$

$$= E(X_1^* X_2^*) + \frac{1}{2} G_1 + \frac{1}{2} G_2$$

and therefore

$$G_{1,2} = \frac{6 \int_{[0,1]^2} \mathbb{L}_{C, L_1^{-1}, L_2^{-1}}^{-1}(u_1, u_2) du_1 du_2 - 1}{5} = \frac{6}{5} E(X_1^* X_2^*) + \frac{3}{5} G_1 + \frac{3}{5} G_2 - \frac{1}{5}.$$

$\square$

It follows from Theorem 4.1 that upper and lower bounds for $G_{1,2}$ are given by

$$\frac{3}{5} G_1 + \frac{3}{5} G_2 - \frac{1}{5} \leq G_{1,2} = \frac{6}{5} E(X_1^* X_2^*) + \frac{3}{5} G_1 + \frac{3}{5} G_2 - \frac{1}{5}$$

$$\leq \frac{6}{5} min\{E(X_1^*), E(X_2^*)\} + \frac{3}{5} G_1 + \frac{3}{5} G_2 - \frac{1}{5}$$

$$= \frac{6}{5} min\{\frac{1}{2} - \frac{1}{2} G_1, \frac{1}{2} - \frac{1}{2} G_2\} + \frac{3}{5} G_1 + \frac{3}{5} G_2 - \frac{1}{5}$$

$$= \frac{6}{5} (\frac{1}{2} - \frac{1}{2} max\{G_1, G_2\}) + \frac{3}{5} G_1 + \frac{3}{5} G_2 - \frac{1}{5}$$

$$= \frac{2}{5} - \frac{3}{5} max\{G_1, G_2\} + \frac{3}{5} G_1 + \frac{3}{5} G_2$$

Note, that the sum of weights is 1 in the lower and upper bound.

Table 1 shows the univariate Gini coefficients and the MEGC for the examples using the Gaussian copula from Figs. 1 and 2.

As expected, the MEGC increases with increasing strength of the dependence between $X_1$ and $X_2$. For the example of wealth and income, the influence of the dependence structure on the MEGC is positive if a rich person tends to belong to the group of high income individuals and is negative if a rich person is more likely to belong to the individuals with low income.

Note, that also values $G_{1,2} > max\{G_1, G_2\}$ and $G_{1,2} < min\{G_1, G_2\}$ are possible to correctly capture the influence of the dependence on the inequality. This is in contrast to a

**Table 1** Univariate Gini coefficients and corresponding MEGC

| Copula type | (Fig. No.) | MEGC |
|---|---|---|
| Gaussian $\rho = -0.9$ | (1a) | 0.25 |
| Gaussian $\rho = -0.5$ | - | 0.29 |
| Gaussian $\rho = 0.0$ | (1b) | 0.33 |
| Gaussian $\rho = 0.5$ | (1c) | 0.39 |
| Gaussian $\rho = 0.9$ | (1d) | 0.43 |
| Gumbel $\rho = 0.8$ | (2a) | 0.42 |
| Clayton $\rho = 0.8$ | (2b) | 0.41 |
| Gaussian $\rho = 0.0$ | (2c) | 0.22 |
| Clayton $\rho = 0.8$ | (2d) | 0.30 |

Values of MEGC for the examples from Figs. 1 and 2. Note that the univariate Gini coefficients are 0.33 in all cases, except for the last two where the marginal Gini coefficient of the first variable equals 0.05. Further notice that the MEGC, unlike a convex combination, is not necessarily enclosed by the marginal univariate Gini coefficients convex combination of $G_1$ and $G_2$. Consider for example the first eight cases in Table 1, where we have $G_1 = G_2 = 1/3$. If the MEGC would be bounded by these values to be equal to 1/3 in all cases, the different dependence structures would not be reflected.

## 5 Nonparametric estimation of the multivariate Lorenz curve (MEILC) and the corresponding multivariate Gini coefficient (MEGC)

We assume that we have observations $X_1, \ldots, X_n$ on $X = (X_1, \ldots, X_d)$, where $X_j = (X_{1j}, X_{2j}, \ldots, X_{dj})$ for $j = 1, \ldots, n$. We only consider the case where $n > d$, where we have more observations than dimensions. If $F_i$ and $L_i$ would be known for $i = 1, \ldots, d$ we could easily derive observations $X_1^*, \ldots, X_n^*$ on $X^*$ with $X_{ij}^* = L_i(F_i(X_{ij}))$ for $i = 1, \ldots, d$ and $j = 1, \ldots, n$.

However, $F_i$ and $L_i$ are unknown in practical applications and have to be estimated using $X_1, \ldots, X_n$.

We estimate $F_i$ for $i = 1, \ldots, d$ by its empirical counterpart

$$\hat{F}_{in}(x) = \frac{1}{n} \sum_{j=1}^{n} \mathbb{1}_{\{X_{ij} \leq x\}} \quad \text{for } x \in \mathbb{R}.$$

$L_i$ is usually estimated by

$$\hat{L}_{in}\left(u = \frac{k}{n}\right) = \frac{\sum_{j=1}^{k} X_{i[j:n]}}{\sum_{j=1}^{n} X_{ij}} \quad \text{for } k = 0, 1, \ldots, n \text{ and }, i = 1 \ldots, d,$$

where $X_{i[1:n]} \leq X_{i[2:n]} \leq \cdots \leq X_{i[n:n]}$ is the increasingly ordered sequence of $X_{ij}$ and linear interpolation between $\hat{L}_{in}(u = \frac{k}{n})$ and $\hat{L}_{in}(u = \frac{k-1}{n})$ for $k = 1, 2, \ldots, n$. This is tantamount to the compact formula

$$\hat{L}_{in}(u_i) = \frac{\int_0^{u_i} \hat{F}_{in}^{-1}(t)dt}{\frac{1}{n} \sum_{j=1}^{n} X_{ij}} \quad \text{for } i = 1, \ldots, d \text{ and } u_i \in [0, 1].$$

It is now possible to estimate the observations $X_{ij}^*$ by what we suggest to call "pseudo-observations" of $X_{ij}^*$ with

$$\hat{X}_{ij,n}^* = \hat{L}_{in}(\hat{F}_{in}(X_{ij})) = \frac{\sum_{l: x_{il} \leq x_{ij}} x_{il}}{\sum_{l=1}^n x_{il}} \quad \text{for } i = 1, \ldots d \text{ and } j = 1, \ldots n \tag{8}$$

and obtain the corresponding vector $\hat{\boldsymbol{X}}_{j,n}^* = (\hat{X}_{1j,n}^*, \hat{X}_{2j,n}^* \ldots, \hat{X}_{dj,n}^*)$ for $j = 1, \ldots, n$.

## 5.1 Estimation of the MEILC

It was pointed out that the MEILC is given by

$$\boldsymbol{u} = (u_1, \ldots, u_d) \longmapsto C\left(L_1^{-1}(u_1), \ldots, L_d^{-1}(u_d)\right) \quad \text{for } \boldsymbol{u} \in [0,1]^d$$

and that it is the joint distribution function of the vector $\boldsymbol{X}^* = (X_1^*, \ldots, X_d^*)$. Therefore, the MEILC is estimated by the empirical distribution function based on $\hat{\boldsymbol{X}}_{j,n}^*$ for $j = 1, \ldots, n$, i.e.,

$$\hat{\mathbb{L}}_{C,L_1^{-1},\ldots,L_d^{-1},n}^{-1}(u_1, \ldots, u_d) = \frac{1}{n}\sum_{j=1}^n \prod_{i=1}^d \mathbb{1}_{\{\hat{X}_{ij,n}^* \leq u_i\}}.$$

## 5.2 Estimation of the multivariate Gini coefficient (MEGC)

In order to estimate $G_{C,L_1^{-1},\ldots,L_d^{-1}}$ we have to estimate the integral

$$I_{C,L_1^{-1},\ldots,L_d^{-1}} = \int_{[0,1]^d} \mathbb{L}_{C,L_1^{-1},\ldots,L_d^{-1}}^{-1}(\boldsymbol{u})d\boldsymbol{u}$$

for $\boldsymbol{u} = (u_1, \ldots, u_d)$, which is usually done by

$$\begin{aligned}
\hat{I}_{C,L_1^{-1},\ldots,L_d^{-1},n} &= \int_{[0,1]^d} \hat{\mathbb{L}}_{C,L_1^{-1},\ldots,L_d^{-1},n}^{-1}(\boldsymbol{u})d\boldsymbol{u} = \frac{1}{n}\sum_{j=1}^n \int_{[0,1]^d} \prod_{i=1}^d \mathbb{1}_{\{\hat{X}_{ij,n}^* \leq u_i\}} du_i \\
&= \frac{1}{n}\sum_{j=1}^n \prod_{i=1}^d \int_0^1 \mathbb{1}_{\{\hat{X}_{ij,n}^* \leq u_i\}} du_i \\
&= \frac{1}{n}\sum_{j=1}^n \prod_{i=1}^d (1 - \hat{X}_{ij,n}^*).
\end{aligned}$$

After normalizing we obtain the estimator

$$\hat{G}_{C,L_1^{-1},\ldots,L_d^{-1},n} = \frac{(d+1)! \frac{1}{n}\sum_{j=1}^n \prod_{i=1}^d (1 - \hat{X}_{ij,n}^*) - 1}{(d+1)! - 1}. \tag{9}$$

# 6 Considerations on transfers

In this section we want to discuss some considerations on transfers and their effect on the MEILC and MEGC. We are aware that this is a very wide and complex topic, so we can not cover it in all its aspects. However, we at least want to share first considerations and encourage further research on this topic.

First, we define the *Correlation Increasing Transformation (CIT)* introduced by Tsui (1998) into the inequality literature and further considered by many authors, e.g. Epstein and Tanny (1980), Atkinson and Bourguignon (1982), Decancq (2012), and Gravel and Moyes (2012) or lately Faure and Gravel (2021).

**Definition 6.1** Correlation Increasing Transformation (CIT)

We are considering two possible distributions or allocations $A$ and $B$ of $d$ variables among a fixed number of individuals. Let, e.g., $t_A$ denote the $d$-dimensional vector of variables of individual $t$ in allocation scenario $A$, with analogue expressions for other individuals and distributions. We say that distribution $B$ is obtained from distribution $A$ by a Correlation Increasing Transformation (CIT), if for two individuals $t$ and $z$ with $d$-dimensional attribute vectors $t_B, z_B \in \mathbb{R}^d$ we have the reallocation

$$t_B := \max(t_A, z_A) \text{ and } z_B := \min(t_A, z_A), \tag{10}$$

while the variable vectors of all other individuals stay unchanged, i.e., $m_B = m_A$ for all other individuals $m \notin \{t, z\}$. Here, max/min denote the element-wise maximum/minimum.

Note, that within our framework, this corresponds to only swapping $\hat{X}^*$ values between two individuals while all others $\hat{X}^*$ values remain unchanged. A distribution $B$ is called a *Correlation Increasing Majorization (CIM)* of distribution $A$ if it is obtained by a finite sequence of CIT's from $A$.

The CIT naturally affects the order of multivariate Lorenz curves (MEILC) as summarized in the following proposition 6.1.

**Proposition 6.1** *Any Correlation Increasing Transformation (CIT) or Correlation Increasing Majorization (CIM) from a distribution $A$ towards a distribution $B$, implies the multivariate Lorenz order $B \succeq A$ from Definition 3.1.*

Keeping in mind, that a CIT only exchanges values of $\hat{X}$ of two individuals (with the same effect to $\hat{X}^*$) and that $\hat{\mathbb{L}}^{-1}$ is the joint distribution function of $\hat{X}^* = (\hat{X}_1^*, \ldots, \hat{X}_d^*)$ for $k = 1, \ldots, d$, the proposition follows directly from Epstein and Tanny (1980).

Consequently, any CIT has also direct implications towards the MEGC.

**Proposition 6.2** *Any Correlation Increasing Transformation (CIT) or Correlation Increasing Majorization (CIM), induces a higher multivariate Gini coefficient MEGC. An analogous statement applies to the decreasing counterparts of the operations, which induce lower MEGCs.*

Proposition 6.2 follows directly from Proposition 6.1 and Definition 4.1.

More intuitively, a CIT does only have an impact on the dependence structure, i.e., the copula, as it only swaps the coupling of realizations in the margins. Then, a CIT results per definition in more concordant dependence structure and consequently to an increased multivariate Gini coefficient.

Further, we want to elaborate the topic with the help of an example motivated by an anonymous referee. We look at a population of three individuals and consider the bivariate case, i.e., $d = 2$, where $X_1$ and $X_2$ might again stand for the individuals' income and wealth, respectively.

|  | | Individual 1 | Individual 2 | Individual 3 |
|---|---|---|---|---|
| Society 1 | $\begin{pmatrix} X_1 \\ X_2 \end{pmatrix}$ | $\begin{pmatrix} 3 \\ 3 \end{pmatrix}$ | $\begin{pmatrix} 4 \\ 4 \end{pmatrix}$ | $\begin{pmatrix} 6 \\ 6 \end{pmatrix}$ |
| Society 2 | $\begin{pmatrix} X_1 \\ X_2 \end{pmatrix}$ | $\begin{pmatrix} 5 \\ 3 \end{pmatrix}$ | $\begin{pmatrix} 2 \\ 4 \end{pmatrix}$ | $\begin{pmatrix} 6 \\ 6 \end{pmatrix}$ |
| Society 3 | $\begin{pmatrix} X_1 \\ X_2 \end{pmatrix}$ | $\begin{pmatrix} 4 \\ 3 \end{pmatrix}$ | $\begin{pmatrix} 3 \\ 4 \end{pmatrix}$ | $\begin{pmatrix} 6 \\ 6 \end{pmatrix}$ |

Society 1 is obtained from Society 3 by a simple CIT between Individual 1 and Individual 2. We therefore expect the MEGC of Society 1 to be higher than that of Society 3. For Society 2 the transfer is more complicated. Society 2 is obtained from Society 1 by a transfer of two units $X_1$ from Individual 2 to Individual 1. Reversely, this is equal to first applying a CIT to Society 2 (increases inequality) and then a transfer of one unit $X_1$ from Individual 2 to Individual 1, which is typically considered to reduce inequality. In this case we can not directly rank the distributions by means of their multivariate inequality from looking at the transfers. However, it is possible to rank the distributions by calculating the MEGC, resulting in $MEGC = 0.121$ for the first society, $MEGC = 0.098$ for the second society and $MEGC = 0.084$ for the third society. As expected, we see that Society 1 is more unequal than Society 3. Furthermore, we now can include Society 2 in the ordering.

More general, Pigou-Dalton transfers are widely known to reduce inequality in the univariate case (Dalton 1920). However, the extension to the multivariate case is not straightforward and multiple suggestions have been made, see e.g. Basili et al. (2017), Bosmans et al. (2009), and Banerjee (2014). The problem at hand in the multivariate case is that both, the marginal distributions and the dependence structure, can be changed at the same time, even in opposite directions, e.g., decreasing inequality in the margins while increasing in the dependence structure. Thus it might not be so clear to define types of pure basis transfers between two individuals which act always in the same directions with respect to margins and dependence structure. Further complicating matters, individuals which are not directly included in the transfer can be effected and general statements are very difficult to make. The following example illustrates the above and hopefully encourages for further research.

*Example 6.1* Multivariate Pigou-Dalton-Bundle-Transfers (PDBT) are defined as nonnegative transfer from one unambiguously richer individual to a poorer individual in each attribute. The amounts or the proportions of the transfers need not be the same for all attributes, i.e., it is possible to transfer only one attribute (see Fleurbaey and Trannoy 2003; de la Vega et al. 2010). Consider the two societies below, where Society 2 is obtained from a PDBT of 1.1 units of $X_1$ from Individual 1 to Individual 4.

|  | | Individual 1 | Individual 2 | Individual 3 | Individual 4 | MEGC |
|---|---|---|---|---|---|---|
| Society 1 | $\begin{pmatrix} X_1 \\ X_2 \end{pmatrix}$ | $\begin{pmatrix} 5 \\ 4 \end{pmatrix}$ | $\begin{pmatrix} 4 \\ 5 \end{pmatrix}$ | $\begin{pmatrix} 3 \\ 2 \end{pmatrix}$ | $\begin{pmatrix} 2 \\ 3 \end{pmatrix}$ | 0.131 |
| Society 2 | $\begin{pmatrix} X_1 \\ X_2 \end{pmatrix}$ | $\begin{pmatrix} 3.9 \\ 4 \end{pmatrix}$ | $\begin{pmatrix} 4 \\ 5 \end{pmatrix}$ | $\begin{pmatrix} 3 \\ 2 \end{pmatrix}$ | $\begin{pmatrix} 3.1 \\ 3 \end{pmatrix}$ | 0.141 |

The transfer from richer Individual 1 to Individual 4 obviously reduces the inequality in $X_1$, but also affects Individuals 2 and 3 leading to another result in the multivariate case. Individual 2 now is the richest in both dimensions whereas Individual 3 is the poorest. Therefore, the transfer increased the dependence, more specific the rank-dependence, within the society which increases inequality in this case leading to a higher MEGC.

Example 6.1 displays the complexity of the topic. Although a transfer seems to reduce inequality at the first glance, it may have an opposing efficacy within the dependence structure of the whole society. In our example, the PDBT results in a clear richest and poorest individual, decreasing the balancing effect of the dependence structure.

For more general considerations on transfers we refer to Epstein and Tanny (1980), Atkinson and Bourguignon (1982), and Decancq (2012), or most lately Faure and Gravel (2021).

# 7 Analysis of Income and Wealth Inequality using the MEILC and the MEGC

In the following section we demonstrate a possible application of the MEILC and the corresponding MEGC. The first example in Section 7.1 uses data for Germany (SOEP 2019) and is implemented in Python 3.8 (Van Rossum and Drake 2009). Section 7.2 is implemented via the LISSY R-Interface (Luxembourg Wealth Study (LWS) Database 2020) and examines the MEGC of 13 additional countries.

## 7.1 MEILC and the MEGC for Germany 2017

We analyse the joint inequality of income and wealth in Germany based on the data provided by the Socio-Economic Panel (SOEP) for 2017 (SOEP 2019). Detailed information about the survey and the methods used in the SOEP are provided by Goebel et al. (2019) and Wagner et al. (2007). The analysis is based on the variable *'i11102'* from the *pequi* dataset for income and *'n0111a'* from the *hwealth* dataset for wealth. Entries with negative values in one of the variables are dropped as suggested by many authors (see e.g., Rehm et al. 2014; Harvey et al. 2017; Saez and Zucman 2016; Formby et al. 1989). The data set refers to households, whereas inequality numbers are usually reported at the individual level. Income and wealth numbers in the data set are therefore broken down to the individual level. To this end, the values (income and wealth) are equivalised with respect of the number of household members by multiplying with $1/(\text{household members})^{0.5}$, see, e.g., OECD (OECD Income (IDD) and Wealth (WDD) Distribution Databases 2017). In a next step, each adjusted pair of income and wealth is replicated $K$ times, where $K$ is the product of the household members (e.g., to get 5 individual entries from a 5 person household) and the integer part of a weight reported in variable *'w11102'*. The variable *'w11102'* corrects for differences in the socio-economic distribution between households in the panel and all households of the country. Additionally, data errors are eliminated by excluding individuals with values lying more than 30 standard deviation off the mean.

The resulting data is presented in Fig. 3. For better visualization, the figure only shows data with income and wealth below EUR 0.1 million and EUR 1 million, respectively. This corresponds to more than 98.0% of the data. The empirical copula of the full data set is shown in Fig. 4 and the resulting MEILC in Fig. 5. For the univariate Gini coefficients we compute $G_1 = 0.29$ for income and $G_2 = 0.65$ for wealth. Thus, inequality in wealth is considerably higher than in income. This difference is also observable in the shape of the margins $L_{income}^{-1}$ and $L_{wealth}^{-1}$ of the MEILC in Fig. 5. The multivariate Gini coefficient of wealth and income yields $G_{1,2} = 0.47$. Due to the moderate positive dependence of wealth and income (Spearman's $\rho$ is $\rho = 0.56$, here) this is slightly higher than it would be for independent variables (compare to Eq. 7 in Section 4).

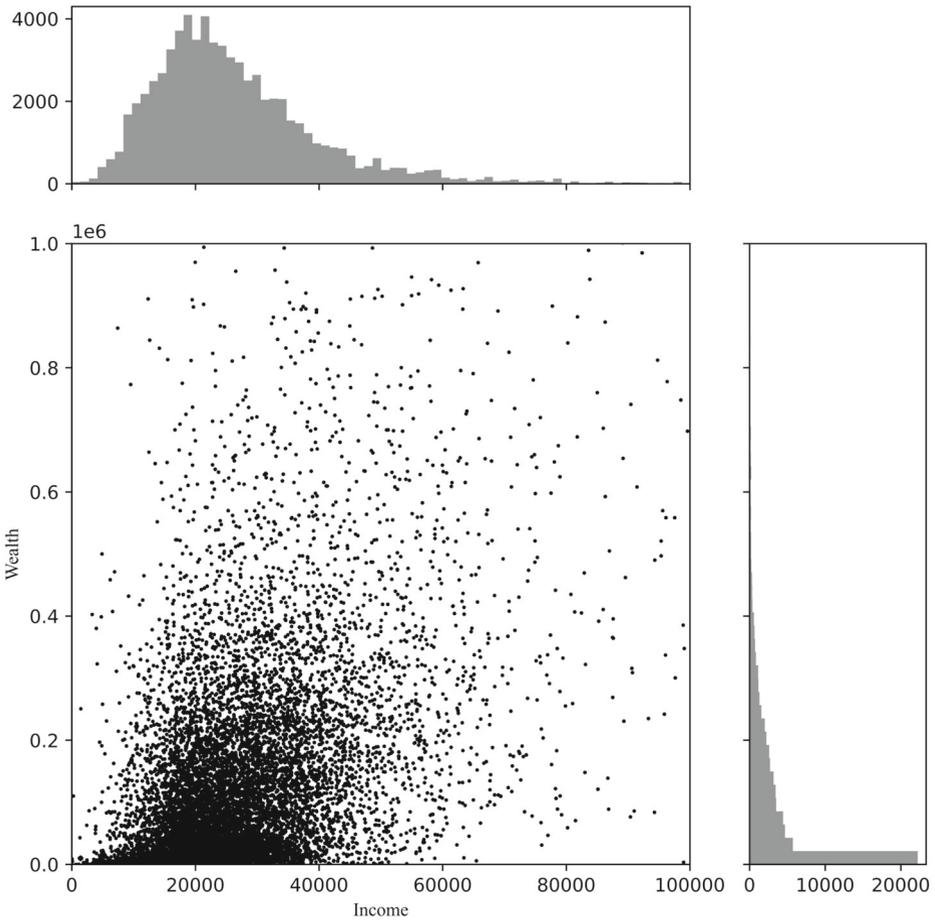

**Fig. 3** Scatter-plot and corresponding histograms of income and wealth in Germany 2017 based on the SOEP dataset

## 7.2 MEGC of income and wealth for other countries

Table 2 summarizes the Gini coefficients for wealth $G_1$, income $G_2$ and the MEGC $G_{1,2}$ for several countries based on the Luxembourg Wealth Study (LWS) database (Luxembourg Wealth Study (LWS) Database 2020). For an extensive documentation of the cross-national wealth database, see LIS (2019a) and LIS (2019b). In the analysis, the variables *disposable household income ('dhi')* and *disposable net worth ('dnw')* from the latest available data sets are used (if too many of the values are missing, we use the *total current income ('hitotal')* variable instead of *'dhi'*). The data processing is done analogously to the SOEP data set

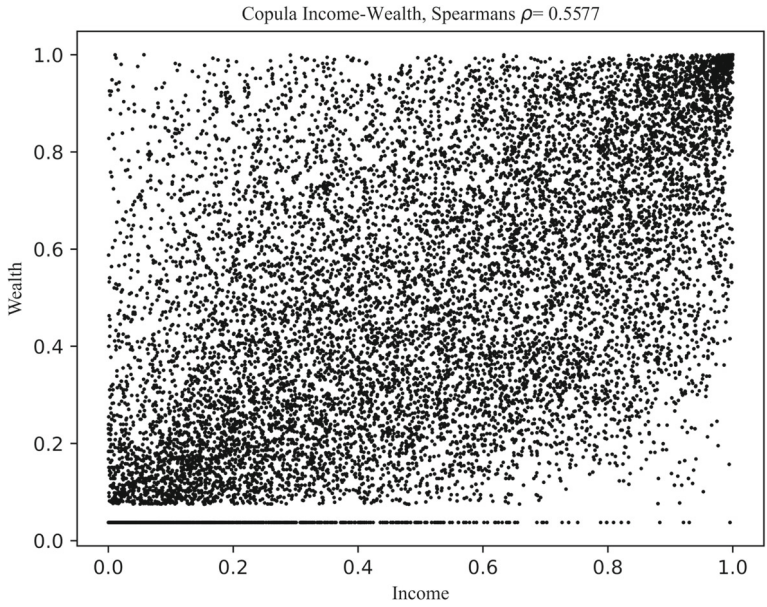

**Fig. 4** Empirical copula of income and wealth in Germany 2017 based on the SOEP dataset

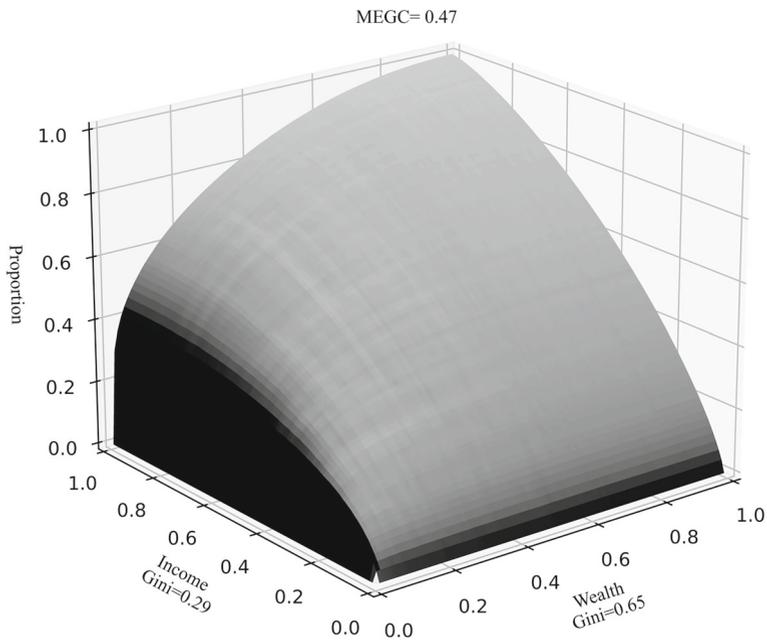

**Fig. 5** MEILC of income and wealth in Germany 2017 based on the SOEP dataset

**Table 2**  Gini coefficient, MEGC and and Spearman's $\rho$ on wealth and income

| Country | Gini on income | Gini on wealth | MEGC | Spearman's $\rho$ |
|---|---|---|---|---|
| Australia | 0.33 | 0.62 | 0.46 | 0.28 |
| Austria[1] | 0.28 | 0.67 | 0.47 | 0.41 |
| Canada | 0.32 | 0.66 | 0.48 | 0.41 |
| Finland | 0.25 | 0.60 | 0.43 | 0.41 |
| Germany[2] | 0.29 | 0.65 | 0.47 | 0.56 |
| Greece[1] | 0.32 | 0.55 | 0.45 | 0.41 |
| Italy | 0.34 | 0.58 | 0.48 | 0.56 |
| Luxembourg[1] | 0.39 | 0.63 | 0.51 | 0.54 |
| Slovakia[1] | 0.34 | 0.51 | 0.44 | 0.40 |
| Slovenia[1] | 0.36 | 0.59 | 0.46 | 0.29 |
| South Africa | 0.61 | 0.85 | 0.71 | 0.43 |
| Spain[1] | 0.38 | 0.60 | 0.50 | 0.45 |
| United Kingdom | 0.35 | 0.60 | 0.48 | 0.55 |
| United States | 0.45 | 0.80 | 0.61 | 0.63 |

Gini coefficient, MEGC and and Spearman's $\rho$ on wealth and income for multiple countries based on the Luxembourg Wealth Study (LWS) Database (2020) database. [1]*hitotal* instead of *'dhi'* from LWS dataset used because of missing values. [2] From SOEP data, Section 7.1 in the last section. Again, households with negative values in one or both variables are excluded. Both household variables are again broken down to individual levels. First we equalize by multiplying by $1/(\text{household members})^{0.5}$ and then we replicate each income and wealth pair in the sample according to the number of household members times the integer part of panel adjustment weights. In this database, the number of household members is stored in the variable *'nhhmem'*, while the adjustment weights are stored in *'hpopwgt'*.

Turning to Table 2, all reported numbers are plausible and inequality is larger for wealth than for income in all cases. With regard to wealth, South Africa as well as the United States have the highest inequality. South Africa shows also the largest inequality in the income distribution. Considering all countries, dependencies between income and wealth are positive and mainly moderate. The reported numbers of Spearman's rho are often below 0.5. For this reason all reported MEGC numbers of multivariate inequality lie well between the univariate Gini coefficients. The highest MEGC is reported for South Africa followed by the United States. The lowest MEGC numbers are reported for Slowakia and Finland. An interesting example for the effect of the dependence structure on inequality is the pair of Italy and Slovenia. While marginal inequalities in wealth and income are slightly lower in Italy than in Slovenia, the stronger dependency of these variables in Italy (i.e., the rich tend to coincide with the higher earners) results in a higher MEGC in Italy than in Slovenia. As an alternative, graphical illustration of the dominance structure within the three aspects, two univariate Gini coefficients and the MEGG, we provide a Hasse-diagram in Fig. 6. There, concordant ordering of all three aspects between two countries, results in a connecting edge within the graph.

Note, that the values for the univariate Gini coefficients can differ from other publications for various reasons. First, we do not apply any top or bottom coding of the data and exclude all individuals with negative values in the variables. Second, we floor the provided weights to the next integer because of computational reasons. Third, we only use complete cases of the datasets. This means we consider a person in our calculation only, if all information (*dhi, dnw, hpopwgt, nhhm*) of the case is available. Furthermore we treated both variables in the same way, especially when adjusting for the household size. See Sierminska and Smeeding (2005) for a brief discussion on the topic. Last, we used only the data provided by the LWS database and did not supplement it with data from other sources.

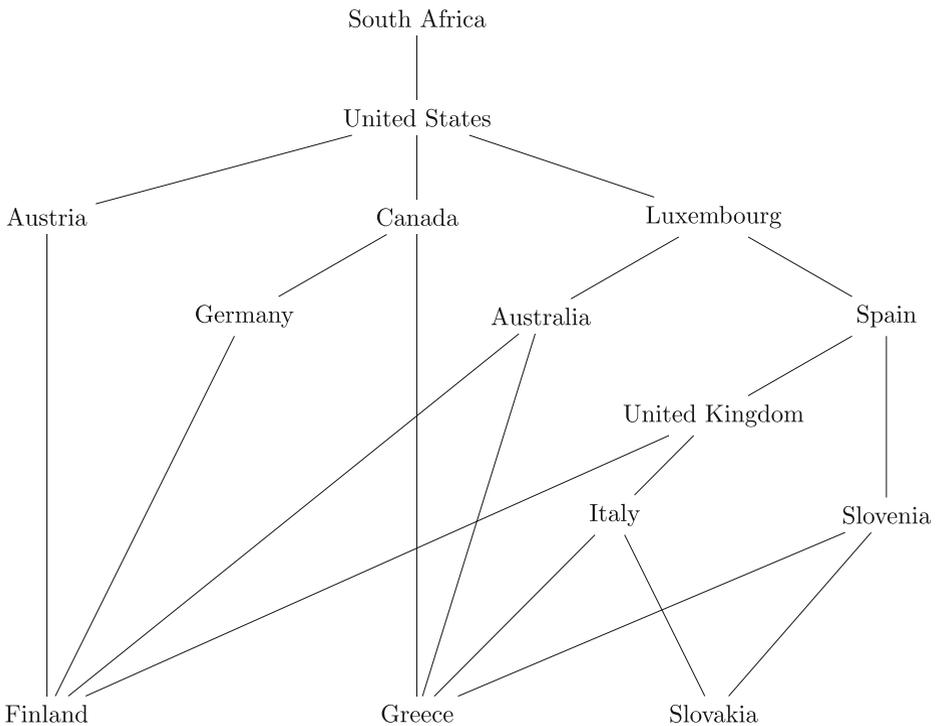

**Fig. 6** Illustration of the incomplete ranking resulting from Table 2 in terms of a Hasse diagram. Dominance in all three Gini coefficients leads to an edge between countries, whereas inequality decreases from the top down

**Funding** Open Access funding enabled and organized by Projekt DEAL.

**Data Availability** Python computer code for the paper is public available in the following repository: https://github.com/FabianKaechele/Multivariate_Extension_Lorenz_Gini Raw datasets analysed in Section 7 are available from (SOEP 2019) and (Luxembourg Wealth Study (LWS) Database 2020) but restrictions apply to the availability of these data, which were used under license for the current study, and so are not publicly available. Information on how to obtain it and reproduce the analysis is available from the corresponding author on request.

## Declarations

**Conflict of Interests** The authors have no relevant financial or non-financial interests to disclose.

and indicate if changes were made. The images or other third party material in this article are included in the article's Creative Commons licence, unless indicated otherwise in a credit line to the material. If material is not included in the article's Creative Commons licence and your intended use is not permitted by statutory regulation or exceeds the permitted use, you will need to obtain permission directly from the copyright holder. To view a copy of this licence, visit http://creativecommons.org/licenses/by/4.0/.

**Publisher's note** Springer Nature remains neutral with regard to jurisdictional claims in published maps and institutional affiliations.